\documentstyle {mn}
\input epsf.sty
\title[Mira PL Relation]
{Globular Clusters and the Mira Period-Luminosity Relation}
\author[M.W.Feast, P.A.Whitelock and J.W.Menzies]
{Michael Feast$^{1}$, Patricia Whitelock$^{2}$ and John Menzies$^{2}$\\
$^{1}$ Astronomy Department, University of Cape Town, Rondebosch, 7701,
South Africa. mwf@artemisia.ast.uct.ac.za\\
$^{2}$ South African Astronomical Observatory, P.O. Box 9, Observatory,
7935, South Africa. paw,jwm@saao.ac.za\\}
\begin{document}

\maketitle

\begin{abstract}
 A globular cluster distance scale based on Hipparcos parallaxes of
subdwarfs has been used to derive estimates of $ M_{K}$ for cluster Miras,
including one in the SMC globular cluster NGC\,121. These lead to a
zero-point of the Mira infrared period-luminosity relation, PL($K$), in good
agreement with that derived from Hipparcos parallaxes of nearby field Miras.
The mean of these two estimates together with data on LMC Miras yields an
LMC distance modulus of 18.60 $\pm $ 0.10 in evident agreement with a
metallicity corrected Cepheid modulus (18.59 $\pm \sim$ 0.10).\\
 The use of luminous AGB stars as extragalactic population indicators is
also discussed.
\end{abstract}


\section{Introduction}
 That Mira variables show a good infrared period-luminosity relation was
established from LMC observations (Glass \& Lloyd Evans 1981, Glass \& Feast
1982, Feast 1984, Feast et al. 1989). In particular Feast et al. (1989)
showed from multi-epoch photometry that the average absolute magnitude 
at $K$ (2.2 $\mu$m) had
a rather small scatter (0.13\,mag) for oxygen-rich Miras from the relation,
\begin{equation}
 M_{K} = -3.47\log P + \gamma.
\end{equation} 
 More recent work (Whitelock \& Feast 2000a and to be published) shows that
in the LMC the bolometric PL relation extends up to at least $\sim$ 1500
days. Menzies \& Whitelock (1985) obtained multi-epoch $JHK$ photometry of a
number of Miras in galactic globular clusters and showed that they could be
fitted to a PL relation, though offset from that of the LMC, on the basis of
the then current distance scale of the clusters and of the LMC. Whitelock et
al. (1994) showed that the LMC Miras and those in galactic globular clusters
fitted the same PL relations if the cluster distance scale were fixed by LMC
globular clusters and an assumed LMC distance. These workers also included
Miras with distances known in other ways (i.e. Miras with companions of known
luminosity). Recently a Mira PL zero-point was obtained from local Miras
with Hipparcos parallaxes (Whitelock \& Feast 2000b). In the present paper
we derive and discuss an independent Mira PL zero-point based on Miras in
globular clusters. This is now possible using a cluster distance scale set
via the Hipparcos parallaxes of subdwarfs (Carretta et al. 2000 and
references there). We also give infrared photometry for the Mira V1 in the
SMC globular cluster NGC\,121 and include it in the analysis. This is the
only Mira with a known period in a globular cluster (and hence with an
estimated metallicity) outside our Galaxy, although long-period carbon Miras
are known in intermediate age Magellanic Cloud clusters (Nishida et al.
2000).

\section{Data} 
The basic data used in this paper are listed in Table 1. 
The $K$ magnitudes listed are 
(except in the case of one star, see section 2.7)
the means of maximum and minimum
of Fourier fits to the available data sets. The latter are
mainly from 
Menzies \& Whitelock
(1985) but referred to Carter (1990) standards (see Appendix). 
In the case of the
three variables in 47 Tuc the multi-epoch photometry of Frogel et al.
(1981) was included. 

The $JHK$ observations of NGC\,121 V1 are given
in Table 2. These were obtained with the MkIII infrared photometer
on the 1.9m reflector at SAAO, Sutherland. 
The uncertainties in these measures are $0.05$\,mag 
except for those on JD 2445949 when they are $\sim$0.08\,mag. 
\begin{table*}
\centering
\caption{Data on Miras in Globular Clusters}
\begin{tabular}{rrlccrcccc}
\multicolumn{1}{c}{Mira}&\multicolumn{1}{c}{[Fe/H]}&
\multicolumn{1}{c}{$ E_{(B-V)}$}&\multicolumn{1}{c}{$V_{\rm ZAHB}$}
&\multicolumn{1}{c}{$ (m-M)_{0}$}&\multicolumn{1}{c}{$K$}&
\multicolumn{1}{c}{$(J-K)$}&\multicolumn{1}{c}{$\log P$}
&\multicolumn{1}{c}{$ M_{K}$}&\multicolumn{1}{c}{note}\\
&&&&&&\\
N104 V1  & --0.67 & 0.055 &     & 13.38 & 6.20 & 1.28 & 2.326 & --7.20 & 1 \\
      V2 &       &       &      &       & 6.30 & 1.16 & 2.307 & --7.05 & \\
      V3 &       &       &      &       & 6.33 & 1.28 & 2.283 & --7.07 & \\
N121 V1  & --1.02 &       &     & 19.09 & 12.39& 0.94 & 2.147 & --6.71 & 2 \\
N5139 V42&       & 0.15  &      & 13.77 & 7.41 & 1.13 & 2.174 & --6.40 & 3 \\
N5927 V3 & --0.46 & 0.47  & 16.72& 14.55 & 7.25 & 1.49 & 2.491 & --7.43 & \\
N6352 L36& --0.64 & 0.21  & 15.30& 13.96 & 7.20 & 1.25 & 2.243 & --6.82 & \\
N6356 V3 & --0.64 & 0.24  &     & 16.20 & 9.02 & 1.30 & 2.342 & --7.24 & 4 \\ 
      V4 &       &       &      &       & 8.99 & 1.36 & 2.316 & --7.27 & \\
      V5 &       &       &      &       & 9.02 & 1.39 & 2.342 & --7.24 & \\
N6553 V4 & --0.50 & 0.63  &     & 14.08 & 6.32 & 1.59 & 2.423 & --7.93 & 5 \\
    $''$  & --0.44 & 0.84  & 16.92& 13.60 &   &       &       & --7.51 & \\
N6637 V4 & --0.68 & 0.17  & 15.95& 14.74 & 7.96 & 1.24 & 2.292 & --6.83 & \\
N6712 V7 & --0.88 & 0.46  & 16.32& 14.25 & 7.44 & 1.36 & 2.280 & --6.93 & \\
N6838 V1 & --0.70 & 0.25  & 14.52& 13.07 & 6.29 & 1.29 & 2.286 & --6.85 & \\
Ter 5 $\rm V_{S}$  &  0.00 & 2.49  &      & 14.17 & 7.73 & 2.30 & 2.356 & --7.11 & 6 \\
      V  &       &       &      &       & 6.62 & 2.44 & 2.389 & --8.22 & \\
 
\end{tabular}
\raggedright

Notes\\
1. distance modulus from subdwarfs (see text). \\
2. $A_{V}$ = 0.10 ;  $V$(HB) = 19.69.\\
3. See text for distance modulus and reddening. \\
4. $V$(HB) = 17.50. \\
5. [Fe/H], $ E_{(B-V)}$, and $V$(HB) from Zoccali et al. (2001b), 
see text. \\
6. [Fe/H], $E_{(B-V)}$, and $V$(HB) from Ortolani et al. (1996), 
see text. \\

\end{table*} 

Carretta et al. (2000) have rediscussed the distances of globular
clusters based on subdwarf parallaxes from Hipparcos. On the basis 
of this they then derive (their equation 3) a relation between
the Zero-Age-Horizontal-Branch (ZAHB) 
absolute magnitude and the cluster metallicity, viz:
\begin{equation}
 {M_{V} \rm (ZAHB) = 0.18 ([Fe/H] + 1.5) + 0.53.} 
\end{equation}
They also derive (their equation 2),
\begin{equation}
{M_{V}\rm (HB) = 0.13 ([Fe/H] + 1.5) + 0.44} 
\end{equation}
for the mean level of the horizontal branch (HB).
Unless otherwise stated we have used equation 2 above 
to derive the distance moduli of the clusters. The values of
$E_{(B-V)}$, [Fe/H] and $ V_{\rm ZAHB}$ used are given in the
table. They were taken from Ferraro et al. (1999). The values of
[Fe/H] are on the scale of Carretta \& Gratton (1997). We have
adopted a value of $ R = A_{V}/E_{(B-V)} =3.1$, (see below). 
The periods are generally those quoted by Menzies \& Whitelock (1985)
from the literature. 
That for L36 in NGC\,6352 is from Whitelock (1986).
The following comments are on specific clusters.

\subsection{NGC\,104 (47 Tuc)}
 The adopted distance modulus is that derived directly from
subdwarf fitting by Carretta et al. (2000). The reddening 
and metallicity are also from that paper. Zoccali et al. (2001a)
derive a distance modulus for 47 Tuc from a white dwarf cooling
sequence which is 0.28\,mag nearer than the one used here. However, 
they indicate that there are a number of problems to be solved
before such a distance is fully reliable. 

\subsection{NGC\,121}
  The distance was derived from the data of Dolphin et al. (2001).
Their adopted metallicity and visual absorption were used together with their
value of $V$(HB) and equation 3, above. 
This would appear to be the correct procedure rather than using
the ZAHB relation as Dolphin et al. did (A. Walker private communication).
The distance modulus finally adopted by Dolphin et al. is
0.13\,mag less than our value. This depends on comparing the 
colour-magnitude diagram
with a theoretical model. In the present paper it seemed best to 
use the HB result to be consistent with the other clusters
and to avoid using theoretical models.
\begin{table}
\centering
\caption{Infrared Observations of NGC\,121 V1}
\begin{tabular}{cccc}
\multicolumn{1}{c}{$\rm \Delta$JD}&{$J$}&{$H$}&{$K$}\\
&&&\\
5577 & 12.90 & 12.24 & 12.09 \\
5603 & 13.19 & 12.38 & 12.20 \\
5604 & 13.22 & 12.33 & 12.15 \\
5621 & 13.54 & 12.55 & 12.40 \\
5647 & 13.70 & 13.06 & 12.75 \\
5652 & 13.72 & 12.97 & 12.80 \\
5688 & 13.42 & 12.37 & 12.64 \\
5949 & 13.47 & 13.04 & 12.70 \\
5953 & 13.47 & 12.84 & 12.64 \\
6042 & 13.40 & 12.58 & 12.34 \\
6090 & 13.37 & 12.73 & 12.49 \\
6100 & 13.34 & 12.68 & 12.54 \\
6113 & 13.10 & 12.33 & 12.22 \\
6394 & 13.18 & 12.38 & 12.19 \\
\end{tabular}
\raggedright

$\rm \Delta$JD = JD--2440000 \\ 
\end{table} 
\subsection{NGC\,5927}
 The cluster membership of the Mira V3 in this cluster has not
been firmly established from radial velocities, although it is near
the centre of the cluster. The period (the longest of any Mira in
a globular cluster) seems well determined (Andrews et al. 1974). 
The distance modulus must be considered uncertain since the
metallicity of the cluster is greater than that of any of the
clusters used to establish
equation 2, the most metal-rich of which is 47 Tuc.
It should be noted, in particular, that there is some evidence
from RR Lyrae variables that the $M_{V}(\rm HB)$ and $M_{V}(\rm ZAHB)$
relations (equations 2 and 3) may steepen at high metallicities
(see e.g. Caputo et al. 2000, and references there).

\subsection{NGC\,6356}
The basic data for this cluster are taken from Bica et al. (1994)
with the cluster metallicity converted to the Carretta-Gratton
scale using equation 7 of Carretta \& Gratton (1997). As in the case
of NGC\,121, equation 3 above has been used together with $V$(HB) to
determine the distance modulus. 

\subsection{NGC\,6553}
 As in the case of NGC\,5927 (above) the metallicity of this cluster
is greater than that of any cluster used to derive equation 2. 
Two distance moduli 
are given. One uses the data on the cluster from Ferraro et al. (1999)
and equation 2 as described above. In the other case the data are taken
from Zoccali et al. (2001b)
and equation 3 is used. Note the considerable difference in
the reddening adopted in the two cases.  
The referee (Dr R. Gratton) has pointed out that a very recent paper
(Carretta et al. 2001) gives $\rm [Fe/H] = -0.06 \pm 0.15$ for this
cluster and that Cohen et al. (1999) preferred $E_{(B-V)} = 0.78$.
With these values and $V{(\rm HB)}$ from Zoccali et al. (2001b)
or $V{(\rm ZAHB)}$ from Ferraro et al. (1999), one obtains
$M_{K} = -7.4$ or $-7.6$  for NGC\,6553 V4. These figures become
$-7.2$ or $-7.4$ if the lower luminosities for high metallicity
RR Lyraes suggested by Caputo et al. (see section 2.3) are used
as a guide. These various values scatter round the value predicted
 by our final Mira PL($K$) relation ($-7.5$). 
\subsection{NGC\,5139 ($\omega$ Cen)}
 The distance modulus of this unusual cluster is quite uncertain at
the present time. We adopt the value used by 
Hughes \& Wallerstein (2000). These
authors say that this distance (and reddening) 
are near the mean of several estimates and also
give the best fit to
isochrones. Since this distance is not derived in the same way
as for the other clusters we do not use $\omega$ Cen in our
estimate of the Mira PL($K$) zero-point. 
A distance modulus of 13.27 has recently been derived 
by van Leeuwen et al. (2000) from a comparison of radial velocities and
proper motions in the cluster.  However, the authors indicate that this
distance should be regarded with considerable caution at the present time. 
 
\subsection{Terzan 5}
The metallicity of this cluster is high, though rather uncertain. As in the
case of the other very metal-rich clusters (NGC\,5927 and NGC\,6553) the
distance determination rests on an extrapolation of the Carretta et al.
relations. In the case of $\rm V_{S}$ there are only observations at two
well separated epochs. However, these differ in phase by about half a period
so the mean of the two (which is given in Table 1) should be a fair
approximation to the mean magnitude. The reddening is very high and varies
across the cluster (Ortolani et al. 1996).  The $(J-K)$ colours of the two
Miras in, or in the direction of, the cluster, together with the
period-colour relation from Feast et al. (1989) show that their reddenings
are similar ($ E_{(J-K)} = 1.10$ for variable $\rm V_{S}$ and 1.22 for
variable V). With a normal reddening law, these values lead to a mean
$E_{(B-V)}$ of 2.37, close to the value adopted by Ortolani et al. for the
cluster.  It is not known whether the Miras are radial velocity members of
the cluster. Since the cluster is projected on the galactic bulge, it is
quite possible that at least one of them is a field star.  The relative
magnitude of the two stars (compare their positions in Fig.\ 1) shows that
if they both lie on the PL($K$) relation, they do not have the same distance
and therefore cannot both be cluster members.

\section{discussion} 
Values of $M_{K}$ are listed in Table 1 as derived from the
data in that table and assuming
$ A_{K} = 0.273 E_{(B-V)}$. 
Figure 1 shows the $ M_{K}$ - $\log P$ plot for these Miras. 
The line is the LMC Mira PL relation (equation 1)
with $\gamma = 0.89$, 
where we have adopted a
Cepheid distance modulus
for the LMC of 18.59. This Cepheid modulus depends on LMC $V,I$ photometry,
a galactic zero-point calibration using parallaxes, proper motions
and radial velocities, pulsation parallaxes and Cepheids in galactic
clusters, together with a metallicity correction (Feast 2001a)  
\footnote{The Cepheid metallicity correction 
for $V,I$ photometry remains somewhat uncertain,
though it seems unlikely to be significantly larger for the LMC than that
used by Feast (2001a). It is of interest to note that the ``Cepheid
galaxies" used by the HST Key project 
(Freedman et al. 2001) to calibrate their value of 
$\rm H_{0}$ have a mean metallicity (weighted according to their
contribution to $\rm H_{0}$) very close to solar. Thus, using a
galactic calibration, 
rather than a calibration based on an assumed LMC distance,
would avoid the need to apply 
any significant metallicity
correction (Feast 2001b).}. 
\begin{figure}
\epsfxsize=8.4cm % fix the y-dimension and scales x-dim. to y-dim.
\epsfbox{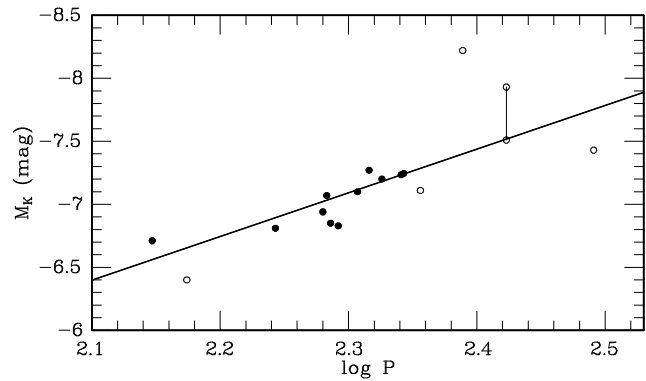}
\caption{The $ M_{K}$ versus $\log$ P diagram for Miras in globular
clusters. The objects with less certain distances are denoted by
open circles. The two points joined by a vertical line are the two
estimates for V4 in NGC\,6553. The line is the LMC Mira relation
with an LMC modulus of 18.59, from Cepheids.}
\end{figure}
 
It is evident in Fig.\ 1 that the agreement in the PL diagram of
LMC field Miras with those in Galactic Globular Clusters and also
the Mira in NGC\,121 (SMC) is good. 
The most discrepant points are for the longer period Miras (V3 in 
NGC\,5927, V4 in NGC\,6553 and, V and $\rm V_{S}$ in Terzan 5). In all
these cases, as already noted, the estimate of the cluster distance
involves an extrapolation
and there are membership, reddening and other uncertainties in 
some of these cases.
Thus the results for these clusters
cannot be given any significant weight in the present discussion. 
As already indicated the distance of $\omega$ Cen
is also still quite uncertain
and determined in a different way to
that of the other clusters. The adopted distance gives a reasonable
fit to the LMC Miras. If the shorter distance of 
van Leeuwen et al. (2001) were
adopted the Mira in the cluster would lie 0.7\,mag 
below the LMC line. Thus if the variable is a normal Mira it provides
some evidence against such a low distance.
 
Omitting the Miras from NGC\,6553, NGC\,5927 and Terzan 5 
and also the Mira in $\omega$ Cen, for the reasons just given,
we are left with 11 Miras in seven clusters. Giving double weight 
to 47 Tuc and NGC\,6356 which each have three Miras and single
weight to each of the other clusters one finds a zero point for
equation 1 of $\gamma = 0.93 \pm 0.14$. The estimated uncertainty 
comes from
the uncertainties given by Carretta et al. (2000) for the coefficients
of equations 2 and 3.
The weighting adopted makes little difference
to the final result. If unit weight is given to each cluster, the
value of $\gamma$ only changes from 0.93 to 0.95. 

The best value to adopt for the ratio, R, of total to selective absorption
is somewhat uncertain. As already noted we adopt 3.1. Some authors
(e.g. Ortolani et al. 2000) prefer to use a larger value (3.3).
Using this larger value would have a significant effect on the results
from clusters with high reddenings. In the case of the very heavily
reddened cluster Terzan 5 the use of $\rm R = 3.3$, would lead to
values of $M_{K}$ which are 0.5\,mag fainter than the values 
given in Table 1 and plotted in Fig.\ 1.
Fortunately the effect is small
for the mean of the clusters used in determining $\gamma$. Adopting
$\rm R= 3.3$ would lead to $\gamma = 0.97$.  

The result for V1 in the SMC globular cluster NGC\,121 is particularly
interesting. 
The star appears to be a normal Mira.
Figure 2 shows the $K$ light curve of this star together with a Fourier
fit. This gives a period of $139 \pm 3$ days in good agreement
with Thackeray's (1958) determination of 140.2 days.
The peak-to-peak pulsation amplitudes derived from Fourier fits
($\Delta J = 0.69$\,mag; $\Delta H = 0.87$\,mag;
$\Delta K = 0.69$\,mag) are within the range shown by globular
cluster and solar neighbourhood Miras of similar period. The mean colours
are $(J-H) = 0.72$\,mag and $(H-K) = 0.20$\,mag. These are bluer
than predicted by the LMC period-colour relations (Glass et al. 1995)
which give 0.83 and 0.25. However, this difference may not be significant. 
The variable has an M-type spectrum with Balmer line emission (Lloyd Evans 1983)
as expected for an oxygen-rich Mira. 
\begin{figure}
\epsfxsize=8.3cm % fix the y-dimension and scales x-dim. to y-dim.
\epsfbox{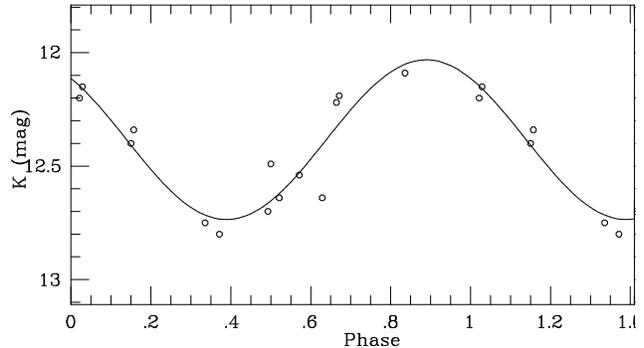}
\caption{Light curve of NGC\,121 V1. The curve
shown is the best fitting sinusoid with a period of 139 d. }
\end{figure}

In a discussion of the
Mira period-metallicity relation, 
Feast \& Whitelock (2000)
found that NGC\,121 V1 fitted the relation well. In that discussion
the globular cluster abundances used were primarily on the Zinn-West (1984)
scale, with [Fe/H] = --1.4 for NGC\,121 being taken from Stryker et al. (1985).
Equation 7 of Carretta \& Gratton (1997) shows that the abundance
(--1.03) adopted above for NGC\,121 from Dolphin et al. (2001), which 
is on the Carretta/Gratton scale, corresponds to --1.23 on the
Zinn-West scale. Changing the metallicity of NGC\,121 from --1.4 to
--1.23 moves the point for V1 in the period-metallicity plot of
Feast \& Whitelock (2000) from just above to just below the mean line
and does not affect the conclusion that this SMC cluster fits the
mean relation well. 

In the present paper we have preferred to use a globular cluster scale based
on Hipparcos parallaxes of subdwarfs rather than rely on theoretical models.
However, it is of interest to compare this scale with that derived by
VandenBerg (2000) from theoretical ZAHB models. For the nine clusters used
by Carretta et al. (2000, table 3) to calibrate equations 2 and 3 (above),
the mean difference between the true moduli derived from their work and
those found from VandenBerg's models is only $0.06 \pm 0.02$, the
VandenBerg scale being shorter. This is despite the fact that VandenBerg's
($\alpha$-enhanced) model comparisons depend on a systematically different
metallicity scale from that adopted by Carretta et al. This suggests that if
the VandenBerg results were used for the clusters containing Miras (with
metallicities placed on his adopted scale), our adopted zero-point which
uses a mean of the cluster result with that obtained from Hipparcos
parallaxes of field Miras (see section 5) would only be made marginally
larger.

\section{AGB Stars as Extragalactic Population Indicators}

In an old or intermediate age population the brightest individual stars will
be on the thermally pulsing asymptotic giant branch (AGB) and the very
brightest of these will be the Mira variables (Feast \& Whitelock 1987;
Whitelock \& Feast 2000a). Because we lack any clear theoretical picture of
evolution at the top of the AGB it has become common practice to compare
extragalactic AGB stars with those in relatively well understood Galactic
environments, e.g. globular clusters or the galactic bulge. Galactic Miras
from specific environments have therefore become calibrators for whole
populations in nearby galaxies wherever individual stars can be isolated and
studied.

Figure 3 shows the position of the globular cluster Miras from Table~1, i.e.
of the tips of the cluster AGBs, in a colour-magnitude diagram which can be
compared with extragalactic systems (after making any appropriate
transformations to the colours). The Miras from Terzan 5 are omitted from
this plot because their membership is far from certain (see section 2.7).
Note that the spread in mean colour is small $\Delta (J-K) \sim 0.4$ for a
range of $\Delta [Fe/H] > 0.5$. A comparison with Davidge's (2000) results
suggests that the spread of age and metallicity in M32 may not be as small
as he suggests.

It is sometimes considered instructive to compare extragalactic populations
with the Galactic bulge and particularly with the M giants (mostly giant
branch rather than AGB stars) in the NGC\,6522 Baade Window (Frogel \&
Whitford 1987). This can be useful, but it is misleading to compare
with the brightest stars in the Baade Window, because the line-of-site depth
and large number of foreground stars in this Bulge field (Feast \& Whitelock
1987; Tiede \& Terndrup 1997) result in some spuriously luminous stars.

Guarnieri et al. (1997a,b) compare the variable stars in NGC\,6553, (see
section 2.5) of which the Mira V4 is the brightest, with AGB stars in M32.
Using a distance modulus of $13.6\pm0.25$ they deduce that
$M_K\sim -8.1$ for V4, which makes it almost as bright as the brightest AGB
stars in M32 (note that all of the data they used, for NGC\,6553 V4 and for
M32 are from single epoch observations). The conclusion drawn is that the
brightest stars in M32 need not be appreciably younger than those in the
globular cluster. Davidge (2001) quotes Guarnieri et al. (1997a) as showing
that V4 has $M_K=-8.5$. In fact V4 has an amplitude of $\Delta K \sim 0.5$
mag and even at maximum light at the larger distance quoted in Table 1 it
will be no brighter than $M_K=-8.2$, if it is at the smaller distance it
will be no brighter than $M_K=-7.8$.
Furthermore the discussion of section 2.5 suggests that the lower of these 
two luminosities is more likely.

Thus whilst the luminosity of the AGB tip as defined by Mira variables
may be a useful population indicator for old systems of known
distance, it must be remembered that these stars are large amplitude
variables. They are unlikely, therefore, to provide very specific
calibrations from single epoch observations.

\begin{figure}
\begin{center}
\epsfxsize=4.5cm % fix the y-dimension and scales x-dim. to y-dim.
\epsfbox{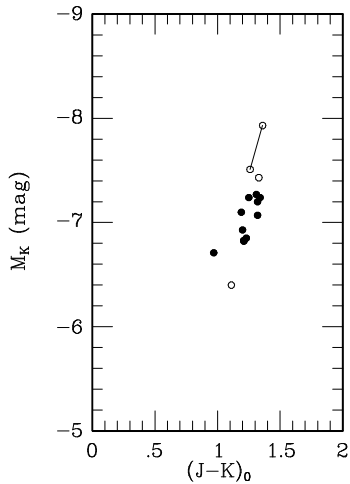}
\caption{ A colour-magnitude diagram for the Miras in Table 1, with symbols
as in Fig.~1 (the Terzan 5 variables are omitted as discussed in the text).}
\end{center}
\end{figure}

\section{Conclusions}
 The zero-point of the Mira PL($K$) relation (equation 1) as derived
above from globular clusters ($0.93\pm 0.14$) agrees satisfactorily
with that obtained from the Hipparcos parallaxes of local (field)
Miras ($0.84\pm 0.14$)(Whitelock \& Feast 2000b). A straight mean
of these two independent estimates is $0.88\pm 0.10$. This zero-point
together with the data on LMC field Miras (Feast et al. 1989) then
yields an LMC true modulus of 18.60, essentially the same as that
derived from a metallicity corrected Cepheid $V,I$ modulus 
($18.59\pm \sim 0.10$, Feast 2001a). 
 
That the Mira in the SMC globular cluster NGC\,121 fits well both
the Mira PL($K$) and $P$-[Fe/H] relations confirms the potential
usefulness of these variables in deriving both distances and
metallicities in old systems including those beyond our own Galaxy. 
Even if periods are not known, Miras, as the brightest AGB stars, can
be used to place some limits on stellar populations in old
extragalactic systems provided mean infrared magnitudes and the
distances of the systems are known.

\subsection*{Acknowledgements}
Dave Laney kindly made a number of the $JHK$ measurements of
NGC\,121 V1. We are grateful to the referee (Raffaele Gratton)
for helpful comments.

\appendix
\section*{Appendix: Note on Photometric Systems}
All the observations used in this paper are on, or have been converted to,
the natural system of the SAAO 1.9m telescope, using the standard
magnitudes from Carter (1990). This is the system adopted by 
Feast et al. (1989) for the data on the LMC Miras. Thus the data of
Menzies \& Whitelock (1985) has been adjusted by the zero-point
corrections specified by Feast et al. (1989, appendix A).
The data from Frogel et al. (1981) have been converted to the 1.9m
system using the tranformations given by Carter (1990) and noting that
$K_{1.9} = K_{c}$ and, for Miras, $(J-K)_{1.9} = 0.955(J-K)_{c}$
(Catchpole, Whitelock, Glass \& Feast, unpublished, see Feast 1996).
Here the subscript ``1.9" refers to the 1.9m natural system and
the subscript ``c" to the ``standard'' SAAO system, as defined by
Carter (1990). 

These conversions
have no significant effect on the PL($K$) results of this paper.
\end{document}